\documentclass[12pt]{iopart}
\usepackage{graphicx}
\usepackage{iopams}
\usepackage{cite}
\newcommand{\be}{\begin{equation}}
\newcommand{\ee}{\end{equation}}

\begin{document}

\title{Modelling on the very large-scale connectome}

\author{G\'eza \'Odor (1), Michael T. Gastner (2), Jeffrey Kelling (3), Gustavo Deco (4)}

\address{(1) Institute of Technical Physics and Materials Science,
Center for Energy Research, P. O. Box 49, H-1525 Budapest, Hungary \\
(2) Yale-NUS College, Division of Science, 16 College Avenue West, \#01-220 Singapore 138527 \\
(3) Department of Information Services and Computing,
Helmholtz-Zentrum Dresden - Rossendorf, P.O.Box 51 01 19, 01314 Dresden, Germany \\
(4)  Center for Brain and Cognition, Theoretical and Computational Group,
Universitat Pompeu Fabra / ICREA, Barcelona, Spain}
\ead{odor@mfa.kfki.hu}
\vspace{10pt}
\begin{indented}
\item[]April 2021
\end{indented}

\begin{abstract}
In this review, we discuss critical dynamics of simple nonequilibrium models
on large connectomes, obtained by diffusion MRI, representing the white
matter of the human brain. In the first chapter, we overview graph theoretical
and topological analysis of these networks, pointing out that universality 
allows selecting a representative network, the KKI-18, which has been 
used for dynamical simulation. The critical and sub-critical behaviour of
simple, two- or three-state threshold models is discussed with special emphasis
on rare-region effects leading to robust Griffiths Phases (GP). 
Numerical results of synchronization phenomena, studied by the Kuramoto model, 
are also shown, leading to a continuous analog of the GP, termed frustrated 
synchronization. The models presented here exhibit dynamical scaling behaviour 
with exponents in agreement with brain experimental data if local homeostasis 
is provided.
\end{abstract}

%
%
%
%
%

\section{Introduction}

The organization of resting-state activity (i.e. \ the dynamics of the brain
that causes switching between different 'functional modes') presumably plays a 
critical role, because it requires a large part of the total energy 
budget~\cite{Atwell01,Raichle06}.
There is empirical and computational evidence showing that the resting organization 
facilitates task-based information processing~\cite{FISER2010119}.
Resting brain networks, as captured by functional connectivity maps,
consistently show that functional connectivity can predict individual 
differences in task-evoked regional activity~\cite{Tavor16,Cole16,Osher19}.
From a mechanistic perspective, whole-brain models can demonstrate
that resting-state activity conforms to a state of criticality that
promotes responsiveness to external stimulation, i.e.\  organization of 
resting-state activity facilitates task-based
processing~\cite{ChialvBak1999_LearningMistakes,
RChial2004_CriticalBrainNetworks,
Chialv2006_OurSensesCritical,
Chialv2007_BrainNear,
ChialvBalenzFraima2008_Brain:What,
FraimaBalenzFossChialv2009_Ising-likeDynamicsLarge-scale,
expert_self-similar_2011,
FraimaChialv2012_WhatKindNoise,
Deco12,66,Senden16}.
We can explore the possible critical dynamics on connectomes, using the methods 
of statistical physics and compare with experiments, but it is unknown what happens 
precisely. There are neuroscience studies, investigating the effects of illness, 
injury or chemicals on the state of brain. However, it was found that personalized whole-brain 
dynamical models poised at criticality track neural dynamics~\cite{Rocha2020RecoveryON}.
Furthermore, all chemically altered states, showed some signs of persistent criticality, 
when exponent relations and universal shape-collapse were tested. 
The maintenance of critical brain dynamics may be important for regulation and control 
of conscious awareness~\cite{Varley2020.03.27.012070}.

Neuronal avalanches are cascading sequences of increasing activations that reveal 
critical behaviour, in which brain functions are optimized by enhancing, for example, 
input sensitivity and dynamic range~\cite{Shew-Plenz-2013}.
This criticality-based optimization is sustained, thanks to the underlying
particular operating regime perched at the brink between phases of order and
disorder. Criticality is explicitly reflected by the scaling invariance of the 
sizes and durations of neuronal avalanches. A signature of criticality is the fact 
that the sizes and durations of neuronal avalanches follow power-law (PL)
statistics with exponents that depend on each other. Furthermore, in the context 
of experimental neuronal avalanches, Friedman et al.~\cite{Fried} demonstrated 
another signature of criticality, namely self-similar scaling of avalanches, by showing that 
the dynamics of long-duration avalanches are similar to those of 
short-duration avalanches when they are properly rescaled. 
Earlier, scaling behavior of avalanches has also been found in signals 
of ruptures, earth-quaqkes and in magnets, like in the random filed
Ising model~\cite{Sethna_2001} and in many systems with phase 
transition to absorbing states~\cite{Munoz99,odorbook}.

On one hand in neuroscience, avalanches are usually considered to be discrete
events, obtained by a temporal binning of LFP signals, introducing a certain degree 
of uncertainty. On the other hand, in models, other than simple branching processes, 
can exhibit continuous order parameter "avalanches", which can also be defined via 
the spreading of growth of the order parameter, if the system is started from an atypical 
state of the control parameters. At the critical point this dynamical behavior is the
so called initial slip scaling phenomena (see for example~\cite{odorbook}),
and the corresponding dynamical exponents can be related to the
exponents of the "avalanches~\cite{Munoz99}. Thresholding of continuous 
signals, to define avalanches, may also introduce some uncertainty.

The criticality hypothesis has been proposed because information processing,
sensitivity, long-range and memory capacity is optimal in the neighbourhood
of criticality~\cite{KC,Chi10,Larr,MArep}.
Criticality in statistical physics is defined by diverging correlation 
lengths and times as we tune a control parameter to the critical value. 
As a consequence, microscopic details are irrelevant, and universal critical
scaling exponents appear in general.

During the last decade, criticality was shown in neuronal recordings (spiking
activity and local field potentials, LFPs) of neural cultures in
vitro~\cite{BP03,Mazzoni-2007,Pasquale-2008,Fried}, LFP signals in 
vivo~\cite{Hahn-2010}, field potentials and functional magnetic
resonance imaging (fMRI) blood-oxygen-level-dependent 
(BOLD) signals in vivo~\cite{Shriki-2013,Tagliazucchi-2012}, voltage 
imaging in vivo~\cite{Scott-2014}, and $10$--$100$ single-unit or 
multi-unit spiking and calcium-imaging activity in 
vivo~\cite{Pris,Bellay-2015,Hahn-2017,Seshadri-2018}.
Furthermore, source reconstructed magneto- and electroencephalographic 
recordings (MEG and EEG), characterizing the dynamics of ongoing cortical activity,
have also shown robust power-law scaling in neuronal long-range temporal 
correlations (LRTC). These are at time scales from seconds to hundreds of seconds
and describe behavioural scaling laws consistent with concurrent 
neuronal avalanches~\cite{brainexp}. However, the measured scaling exponents
do not seem to be universal.
Instead, they are scattered around the mean-field values of 
the Directed Percolation (DP) universality class~\cite{odorbook}. 
It is worth mentioning that LRTC allow investigating scaling at
different time scales than avalanche measurements, which span only 
the $0.001$--$0.1$ secundum range.
Similarly, recent calcium imaging recordings of dissociated neuronal cultures 
show that the exponents are not universal, and significantly different 
exponents arise with different culture preparations~\cite{Yag}.
Very recently it has been found that in dissociated cortex cultures, which lack 
the differentiation into cortical layers, a first-order phase transition
is expected not following the universal mean-field exponents~\cite{Plenz21soc}.
Thus, the single mean-field universality class hypothesis has been challenged
and different explanations have been suggested, ranging from subsampling of 
the branching DP model~\cite{Carval21} to external source effects~\cite{Fosque-20}. 
External sources, leading to the well known Widom line phenomena have been studied 
both by experiments and simulations. Quasicriticality, generated by external excitations,
was suggested to explain the lack of universality observed in
different experiments~\cite{Fosque-20}.
However, it was discovered earlier that heterogeneity can cause non-universal
exponents as a consequence of rare regions and Griffiths 
Phases in brain models~\cite{MM,HMNcikk,GS16,GS18}.

Nevertheless, at the whole-brain level, criticality remains still an open
question.  Mesoscopic local measurements (e.g.\ LFP or spiking activity) might
introduce a bias due to the limited number of neurons and, consequently, are
sensitive to subsampling effects~\cite{Stumpf-2005,Pris,Levina-2017,Carval21}. 
Furthermore, in electrode experiments, there is uncertainty about partitioning 
and binning of activity spots into avalanches.
Earlier, temporal closeness of measured spikes was used to determine
the beginning and the end of activity spikes~\cite{beggs_neuronal_2003}, 
but later this method was criticized~\cite{priseman_critic-14},
and data analysis has become more careful.
However, BOLD or MEG signals might not capture the global dynamics
if the parcellation used is too coarse. Therefore, to study criticality in the nervous
system, it is necessary to monitor whole-brain dynamics with high resolution.

Until now, structural network studies were performed on much smaller-sized
connectomes. For example, the data obtained by Sporns and collaborators, using
diffusion imaging techniques~\cite{hagmann_mapping_2008,honey_predicting_2009}, 
consist of a highly coarse-grained mapping of anatomical connections 
in the human brain, comprising $N = 998$ cortical areas and the fiber 
tract densities between them.  
On the other hand, connectome-based epidemic models may also bring us information 
about a particular disease, for clinical applications see~\cite{GS-epi}.

In order to prove real power laws, valid for several decades, one needs to 
consider larger-sized systems, lack of corrections to scaling and size cutoff~\cite{MM}.
For example the autocorrelation function near to a critical point is of the form: 
\begin{equation} 
A(t) = ( A_0 t^{-\Lambda} + A_1 t^{-\Lambda_1} + ... ) F[ \exp{(-t/\tau)} ]  \ , 
\end{equation}
thus besides the leading order PL singularity $A_0 t^{-\Lambda}$
faster decaying, sub-leading scaling correction terms $A_i t^{-\Lambda_i}$
and an exponential cutoff can occur at large times.

Therefore, we have downloaded the largest available human connectomes and 
run numerical simulations to test the criticality hypothesis, with a 
focus on heterogeneity effects and the possibility of emergence of dynamical
criticality sub-critically.
Since this graph has a graph (also called topological) dimension 
$d < 4$~\cite{gastner_topology_2016}, a real synchronization phase 
transition is not possible in the thermodynamic limit~\cite{HPClett}.
Still, we could locate a transition between partially synchronized and 
de-synchronized states.
In this review, we show simulation results based on direct measurements of 
characteristic times and sizes instead of using avalanche spot binning.

It has been debated how a neural system is tuned to criticality.
At first, self-regulatory mechanisms \cite{stas-bak}, leading to
self-organized criticality \cite{pruessner}, were proposed.
Let us remark that, as a consequence of heterogeneity, extended dynamical 
critical regions with non-universal scaling emerge naturally in spreading 
models~\cite{MM,HMNcikk,GS16,GS18}.
If quenched heterogeneity (i.e.\ disorder with respect to a homogeneous 
system) is present, rare-region effects~\cite{Vojta2006b} and an extended 
{\it semi-critical} region where spatial correlations do not diverge,
known as Griffiths Phase~(GP)~\cite{Griffiths}, can emerge.
Rare regions are very slowly relaxing domains that remain, for a long time, 
in the phase that is the opposite of the phase of the whole system, 
causing slow evolution of the order parameter.
In the entire GP, which is an extended control parameter region around the
critical point, the susceptibility diverges, providing a high sensitivity 
to stimuli, which is beneficial for information processing.
Auto-correlations follow fat-tailed power-laws, resulting in 
burstiness~\cite{burstcikk}, which is a frequently observed phenomenon
in human behaviour~\cite{Karsai_2018}.
Even in infinite-dimensional systems, where mean-field behaviour is expected,
Griffiths effects~\cite{Cota2016} may occur in finite time windows.
As real systems are mostly inhomogeneous, one must assess whether the
heterogeneity is weak enough to justify the usage of homogeneous models 
for describing them. Also one must asses if the heterogeneity can be
considered quasistatic with respect to the time scale of the neuronal activity. 
It was also proposed that a GP might be the reason for the working memory in the 
brain~\cite{Johnson}.  

As individual neurons in-vitro emit periodic signals~\cite{PSM16},
it is tempting to use oscillator models and to investigate criticality
at the synchronization transition point. Recently, a brain model analysis 
using Ginzburg--Landau type equations concluded that empirically reported
scale-invariant avalanches can possibly arise if the cortex is operated at 
the edge of a synchronization phase transition, where neuronal avalanches 
and incipient oscillations coexist \cite{MunPNAS}.
Several oscillator models have been used in biology.
The simplest one is the Hopf model~\cite{Freyer6353}, which has
been used frequently in neuroscience because it can describe
a critical point with scale-free avalanches that have a sharpened
frequency response and enhanced input sensitivity.

Another complex model, describing more non-linearity, is the Kuramoto 
model~\cite{kura,Acebron} that was studied analytically and computationally
in the absence of frequency heterogeneity on a human connectome graph 
with $998$ nodes and on hierarchical modular networks, 
in which moduli exist within moduli in a nested way at various 
scales~\cite{Frus}.
Because of quenched, purely topological heterogeneity,
an intermediate phase was found between the standard synchronous and
asynchronous phases, showing ``frustrated synchronization'',
meta-stability and chimera-like states~\cite{chimera}.
This complex phase was investigated further in the presence of
noise~\cite{Frus-noise} and on a simplicial complex model of manifolds
with finite and tunable spectral dimension \cite{FrusB} as a simple model
for the brain. 

The dynamical behaviour of the heterogeneous Kuramoto model, especially for local
interactions, is a largely unexplored field to the best of our knowledge.
In case of identical oscillators, heterogeneous phase lags or couplings
have been shown to result in partial synchronization and stable chimera
states~\cite{Feng_2015,PhysRevLett.101.264103,LAING20091569,PhysRevE.89.022914}.
Realistic models of the brain, however, require oscillators~\cite{CABRAL2011130}
to be heterogeneous.
Consequently, one of the main focuses of this review is to discuss criticality  
at the whole-brain level from the perspective of synchronization in the 
heterogeneous Kuramoto model. We also compare the emerging dynamaical behavior 
with those of a discrete threshold models on the same large human connectomes.

\section{Human connectome topology}

The connectome is defined as the structural network of neural
connections in the brain~\cite{sporns_human_2005}.
The human brain has $\approx 10^{11}$ neurons, which current imaging
techniques cannot comprehensively resolve at the scale
of single neurons.
Studies must, therefore, work with coarse-grained data.
Early studies, based on 998 large-scale cortical regions, gave some
insights into the relation between structural and functional
connectivity~\cite{hagmann_mapping_2008, honey_predicting_2009}.
They found that, although strong functional connections exist between
regions that have no direct structural connection, resting state
functional connectivity is constrained by the connectome structure.
Here, we review results about the human connectome at a more
fine-grained level ($\approx 10^{6}$ nodes).
The results are based on diffusion tensor imaging data by Landman et
al.~\cite{landman_multi-parametric_2011}.
Diffusion tensor imaging has generally been found to be in good
agreement with ground-truth data from histological tract
tracing~\cite{delettre_comparison_2019}.
Inferred networks of structural connections were made available by
the Open Connectome Project and previously analyzed by Gastner and
\'Odor~\cite{gastner_topology_2016}.
These graphs are symmetric, weighted networks, where the weights 
measure the number of fiber tracts between nodes.

\subsection{Degree distribution}

Since the early days of complex network science, it has often been
hypothesized that structural and functional brain networks have
power-law degree distributions~\cite{eguiluz_scale-free_2005,
  van_den_heuvel_small-world_2008, kaiser_tutorial_2011}.
In statistical physics, power laws occur at transitions between
ordered and disordered phases.
Because circumstantial evidence supported the hypothesis that the
brain operates near such a critical point~\cite{beggs_neuronal_2003,
  pasquale_self-organization_2008, expert_self-similar_2011}, a
power-law degree distribution appeared to be a plausible assumption.
However, observational data of structural brain networks did not
support the power-law hypothesis for the degree 
distribution~\cite{humphries_brainstem_2006,ivkovic_statistics_2012,
allard_navigable_2020}.

In~\cite{gastner_topology_2016}, we applied model selection based on
the Akaike information criterion to determine which probability
distributions fit the Open Connectome data best.
The investigated candidate distributions for model selection were
power-law, exponential, log-normal and Weibull distributions.
In the case of the power-law and Weibull distributions, we also
considered three-parameter generalizations: a truncated power law,
$P({\rm degree} \geq k) \propto \alpha^\beta (k + \alpha)^{-\beta}
e^{-\gamma k}$, and a generalized Weibull 
distribution~\cite{teimouri_three-parameter_2013}, $P({\rm degree}
\geq k) \propto \exp[\alpha(\gamma^\beta - (k + \gamma)^\beta)]$,
where $k$ is the degree, and $\alpha$, $\beta$ and $\gamma$ are
parameters to be fitted to data.
We also allowed each candidate model to be valid only for sufficiently
large degree $k$ because corrections may, in practice, need to be
applied to small degrees.
The Akaike information criterion imposes a penalty for every
additional parameter in the model that must be fitted to data.
In ~\cite{gastner_topology_2016}, we applied the criterion by Burnham and 
Anderson~\cite{burnham_model_2002} and ~\cite{gastner_topology_2016} 
that there is only empirical support for a statistical model if the 
corresponding Akaike information criterion differs by less than 10 
from the minimum for all candidate models.

In~\cite{gastner_topology_2016}, we fitted parameters to ten human 
connectomes and found that, in nine cases, there was at least some 
empirical support for a generalized Weibull distribution.
In seven cases, there was a similar level of evidence for a truncated
power law.
There was little or no empirical support for any other candidate
model, which excludes power law tails from the set of plausible
assumptions.
Given empirical brain data, the Akaike information criterion generally
favored models with a larger number of parameters, which implies that
it is difficult to mimic the true complexity of the connectome
topology with simple models.
Therefore, we recommend that simulations of brain dynamics are
performed on empirical, rather than modeled, networks if possible.

\subsection{Small-world topology}

Apart from the degree distribution, dynamic processes on networks
also depend on the paths between nodes.
A common feature of many real-world complex networks is their
small-world topology~\cite{watts_collective_1998}.
A network is called a small world if there is, on one hand, a tendency
for nodes to form local clusters, but, on the other hand, there are
also long-distance links that significantly reduce the average of the
shortest path lengths $L$ between all pairs of nodes.
The large-scale network backbone of the brain regions is a small
world~\cite{vaessen_effect_2010}, but it remains an open question
whether the connectome is also a small-world network at a cellular
level~\cite{hilgetag_is_2016}.

To shed light on this issue, we quantified the small-world coefficient
$\sigma$, defined in~\cite{humphries_network_2008}, for Open Connectome networks~\cite{gastner_topology_2016}.
The small-world coefficient compares the clustering coefficient $C$
and the average path length $L$ with the corresponding values $C_r$
and $L_r$ of an Erd\H{o}s--R\'enyi random graph according to the
formula
\begin{equation}
  \sigma = \frac{C / C_r}{L / L_r}\ .
\end{equation}
By definition, a small-world network satisfies $\sigma > 1$.
For the Open Connectome data, we found that $\sigma$ is in the range
from $750$ to $890$ if the clustering coefficient is measured with the
Watts-Strogatz formula~\cite{watts_collective_1998}.
We found no significant dependence of $\sigma$ on the number of nodes
(as shown in Table 4 of ~\cite{gastner_topology_2016}).
Thus, connectomes are small-world networks, and $\sigma$ can be
treated as scale-independent.

\subsection{Graph (topological) dimension}

The small-world coefficient $\sigma$ captures the relation between
local clustering and the mean topological distance (i.e.\ the average
of the minimum number of edges on a path between two nodes).
Another measure of interest is the so called topological or graph dimension 
$D$~\cite{NW99}, which characterizes how quickly neighbourhoods increase as 
a function of distance. Here the distance $r$ is the number
of egdes along the shortest path connecting two nodes.
If $N_r$ is the number of node pairs within a distance $\leq r$, then
the topological dimension $D$ is the exponent that fits best to the
relation $N_r \propto r^D$.
It has been conjectured that a finite dimension $D$ is indicative of
Griffiths phases and rare-region effects, which might explain
power-laws and scale invariance observed in brain network
dynamics~\cite{moretti_griffiths_2013}.

\begin{figure}[h]
  \centering
\includegraphics[width=.5\textwidth]{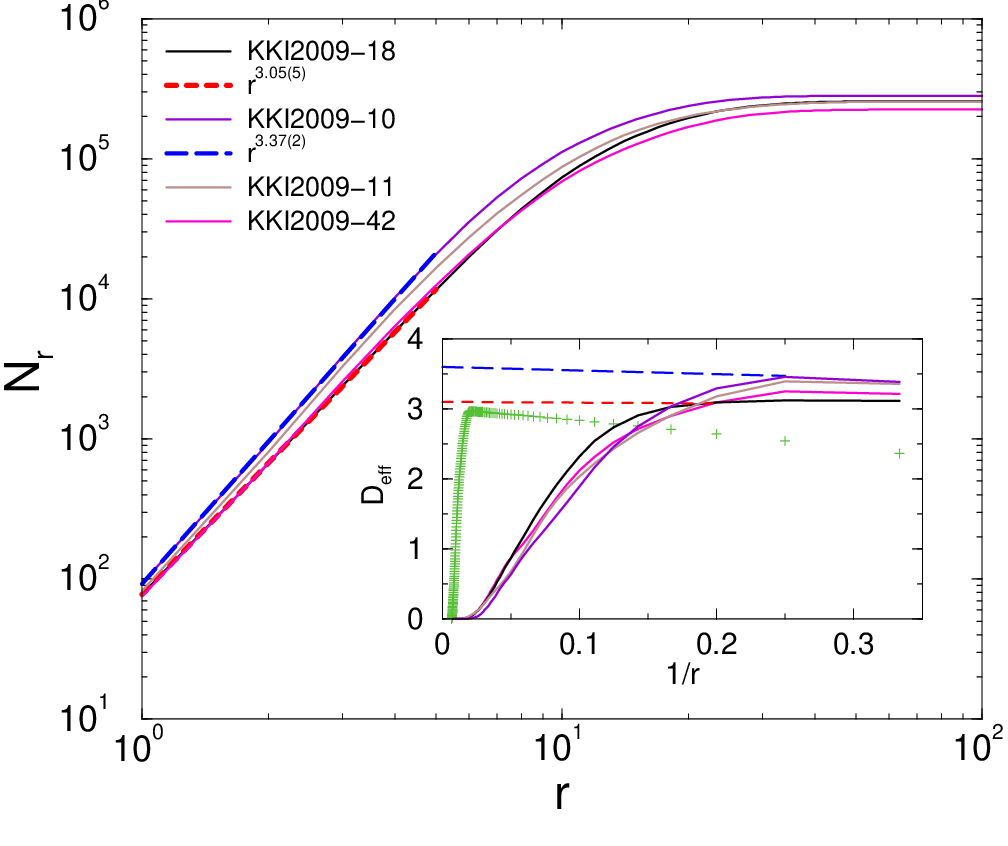}
\caption{\label{dim-KKI}Number of nodes within graph distance $r$
in four different KKI connectomes. Dashed lines show power-law
fits. Inset: local slopes defined in Eq.~\ref{Deff}.
Crosses correspond to measurements on a regular $100^3$ lattice.
Figure from~\cite{gastner_topology_2016}.}
\end{figure}

For the Open Connectome data, power-law fits in the range $1\leq r
\leq 5$ (Fig.~\ref{dim-KKI}) suggest topological dimensions between
$D=3$ and $D=4$~\cite{gastner_topology_2016}.
For larger $r$, $N_r$ saturates because the number of nodes in the
network is finite, so the range over which $\log(N_r)$ increases
linearly in $\log(r)$ is small.
In the inset of Fig.~\ref{dim-KKI}, we plot the discretized derivative
\begin{equation}  \label{Deff}
D_\mathrm{eff}(r+1/2) = \frac {\ln N_r - \ln N_{r+1}} {\ln(r) - \ln(r+1)} \ .
\end{equation}
We estimated $D$ in the limit of large network size by extrapolating
$D_\mathrm{eff}$ (dashed lines in Fig.~\ref{dim-KKI}).
We detected a tendency that larger networks have larger values of
$D$.
However, the increase is small so that Griffith phases in brain
dynamics are plausible.

\subsection{Universality and homeostasis}

The network topology study exhibits a certain level of universality in the 
topological features of the 10 large human connectomes investigated:
degree distributions, graph dimensions, clustering and small world coefficients. 
These can be observed in Tables 3 and 4 of~\cite{gastner_topology_2016}.
It is outside the scope of this review, to compare our previous results with 
the results of other brain studies. 
Instead, we use our previous results in the forthcoming simulations.
One may expect the same dynamical behaviour to occur in dynamical simulations 
on human connectome graphs. 
Therefore, one of the graphs, called KKI-18, was selected
to be the representative in further studies. 
The graphs, downloaded in 2015 from the Open Connectome project 
repository~\cite{OCP}, was generated via the MIGRAINE pipeline~\cite{MIG},
publicly available from~\cite{m2g}.
It comprises a large component with $N = 804\,092$ nodes connected via 
$41\,523\,908$ undirected edges and several small disconnected sub-components, 
which were ignored in the modeling.
The large number of nodes is because of other parcellations closer to 
voxel resolution being used. For instance, there are approximaily 1.8 million 
voxels in the brain mask of a 1 mm resolution standard-aligned MRI. 

\begin{figure}[h]
\centering
\includegraphics[width=.6\textwidth]{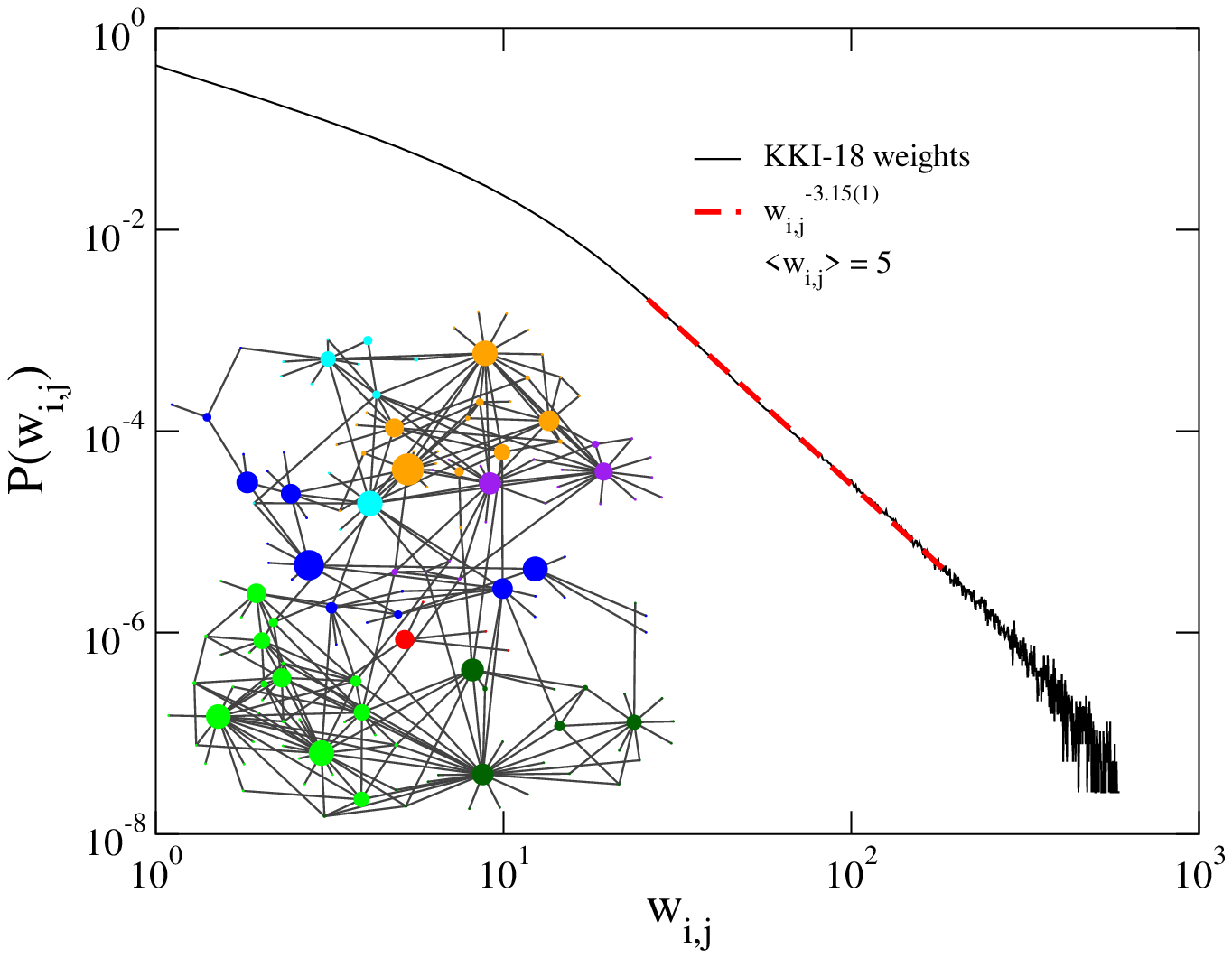}
\caption{\label{pAw} Probability distribution function of link weights in the KKI-18 connectome. 
Dashed line: a power-law fit for intermediate weights. Inset: network of the modules 
of the KKI-18 human connectome graph obtained through Leiden community
 analysis. The areas of the circles is proportional 
to the number of nodes. Each circle's color indicates its membership in
one of seven modules of the displayed graph of modules, illustrating the
 hierarchical nature of KKI-18.}
\end{figure}

The KKI-18 graph exhibits hierarchical modular structure, because it is constructed 
from cerebral regions of the Desikan-Killany-Tourville Parcellation, 
which is standard in neuroimaging~\cite{DESIKAN2006968,10.3389/fnins.2012.00171} 
providing (at least) two different scales.
The graph topology is shown in Fig.~\ref{pAw}, in which modules were 
identified by the Leiden
algorithm~\cite{TraagWaltmaVanEck2018_LouvainLeiden}.
It found $153$ modules, with sizes varying between $7$ and $35\,332$ nodes.
Note, that state of the art tractography results in a structural connectome,
representing real white-matter tracts of a little more than 1000 ROI's; see 
for example~\cite{Misic,zhang2021quantitative,https://doi.org/10.1002/mrm.20642}.

Weights between nodes $i$ and $j$ of this graph vary between $1$ and $854$.
The probability density function is shown in Fig.~\ref{pAw}.
Following a sharp drop, one can observe a power-law region for
$20 < w_{ij} < 200$ with cutoff at large weights. The average weight
of the links is $\simeq 5$. Note that the average degree of this graph 
is $\langle k \rangle=156$~\cite{gastner_topology_2016},
whereas the average number of the incoming weights of nodes is
$\langle W_i\rangle = 1 / N \sum_i\sum_j w_{ij} = 448$.

In Refs.~\cite{CCdyncikk,KurCC}, it was shown that dynamical models,
running on the KKI-18 graph do not exhibit critical transition.
Due to the large weight differences and graph dimensionality, the hubs
cause discontinuous transitions. Nodes with high degrees/weights 
influence many nodes with low degrees/weights. If their state is 
changed the whole systems turns into another state, causing a jump
or jumps in the order parameter, thus the transition will be discontinuous.

However, it was found that certain degree of local homeostasis occur 
in real brain via inhibitory neurons~\cite{Homeo-inh,65,66,67,68}.
This has been modeled by normalizing the incoming interaction 
strengths~\cite{CCdyncikk}. To keep the local sustained activity 
requirement for the brain~\cite{KH} and to avoid nodes, which practically 
cannot affect the activity propagation, the incoming weights were
renormalized by their sum:
\begin{equation}
W'_{ij} = W_{ij}/\sum_{j\; \in\; \rm{neighb.\ of} \ i} W_{ij} \ .
\end{equation}
This renormalization makes the system locally homeostatic, and simulations 
have showed the occurrence of criticality as well as Griffiths effects.
Indeed, there is some evidence that neurons have a certain adaptation 
to their input excitation levels \cite{neuroadap} and can be modeled 
by variable thresholds \cite{thres}.
Recent theoretical studies have also suggested that homeostatic
plasticity mechanisms may play a role in facilitating
criticality \cite{65,66,67,68}.
Recently, a comparison of modelling and experiments arrived at
a similar conclusion: equalized network sensitivity improves the
predictive power of a model at criticality in agreement with
the fMRI correlations~\cite{Rocha2008}.
It is important to realize, that depending on the conditions 
(relative threshold, inhibitory interactions, ... etc.), models 
on the human KKI-18 graph can describe first, second or perhaps
mixed order transitions.

\section{Critical dynamics of discrete threshold models on the connectome}
\label{dec:CCD}

To understand the collective behaviour of a large amount of neurons,
simple discrete models have been used in computer simulations~\cite{KH,Hai}.
The KKI-18 connectome was first studied using a two-state threshold model,
in which sites could be inactive or active: $x_i = 0 \ {\rm or} \ 1$~\cite{CCdyncikk}. 
Later, this study was extended to a three-state model~\cite{GProbcikk}
in which sites could become refractory ($x_i = -1$) for one time step,
when the activity of the state was lost, before becoming inactive.
This prevents activated neighbours from immediately reactivating the source,
which can lead to propagating fronts, resembling to dynamical percolation~\cite{rmp}.
However, this critical behaviour can occur only in the infinitely long 
refractory state limit. 
The application of refractory states is a common feature in brain 
modelling~\cite{Kandel}. It was used in the pioneering brain model
by~\cite{Hai}, as well as in a recent numerical work~\cite{Rocha2008}.
These studies obtained critical behavior, but due to the 998 node
sized connectome used they could not resolve the heterogeneity effects 
and found a Griffiths Phase. Note, that ~\cite{Rocha2008} applied the
same weight normalization as ~\cite{CCdyncikk}.

Dynamical processes, mimicking neuronal avalanches, were initiated by
activating a randomly selected 'seed' node. This was done once at the
beginning of the time loop of the simulations. 
At each network update, every node $(i)$ was visited and tested if the
sum of incoming weights $W_{ij}$ of active neighbours reached a given
threshold value
\begin{equation}
\sum_{j} \delta(x_j,1) W_{ij}  > K \ ,
\end{equation}
where $\delta(i,j)$ is the Kronecker delta function.
If this condition was met, activation of $x_i = 0$ was attempted
with probability $\lambda$. Alternatively, an active node was
deactivated with probability $\nu$. For the refractory version,
the intermediate state ($x_i = 1 \to x_i = -1$) was set with 
probability 1 and deactivation to $x_i = 0$ was done only
at the following graph update.

New states of the nodes were overwritten only after a full network
update.
Until then, they were stored in a temporary state vector.
This procedure means that synchronous updates were performed at discrete time
steps. The updating process continued as long as active sites 
were available or up to a maximum time limit of $t = 10^6$ to $10^7$ 
Monte Carlo sweeps.
For this stochastic cellular automaton, synchronous updating
is not expected to affect the dynamical scaling behaviour~\cite{HMNcikk}
because there are no activity currents.
Synchronous updating makes it possible to implement parallel algorithms.
In fact, the code was implemented on GPUs, which resulted in a $~ 12\times$ speedup
with respect to contemporary CPU cores.

In case the system had fallen into the inactive state, the actual 
time step was recorded in order to calculate the survival probability 
$p(t)$ of runs. The average activity
\begin{equation}
\rho(t) = 1/N \sum_{i=1}^N  \delta(x_i,1)
\end{equation}
and the total number of activated nodes
\begin{equation}
s = \sum_{i=1}^N \sum_{t=1}^T \delta(x_i,1)
\end{equation}
during the avalanche of duration $T$ was calculated at the end of 
the simulations.

By varying the control parameters ($K$, $\lambda$ and $\nu$), one can
locate the transition point between active and absorbing steady states. 
At critical phase transition points, the avalanche survival 
probability is expected to scale asymptotically as
\begin{equation}\label{Pscal}
P(t) \propto t^{-\delta} \ ,
\end{equation}
where $\delta$ is the survival probability exponent \cite{GrasTor}.
This scaling law is connected to the avalanche-duration scaling
\begin{equation}
p(t) \propto t^{-\tau_t}\ .
\end{equation}
Integration in time imposes the exponent relation
\begin{equation}\label{tau-del}
\tau_t = 1+\delta.
\end{equation}
In seed simulations, the number of active sites initially grows as
\begin{equation}\label{Nscal}
N(t) \propto t^{\eta} \ ,
\end{equation}
with the exponent $\eta$, related to the avalanche size distribution
\begin{equation}
p(s) \propto s^{-\tau} ,
\end{equation}
via the exponent relation~\cite{MAval}
\begin{equation}\label{tau-eta}
\tau=(1+\eta+2\delta)/(1+\eta+\delta) \ .
\end{equation}
To analyze corrections to scaling, one can calculate the local slopes
of the dynamical exponents $\delta$ and $\eta$ as the discretized,
logarithmic derivatives of (\ref{Pscal}) and (\ref{Nscal}).
For example, the effective exponent of $\delta$ is measured as
\begin{equation}  \label{deff}
\delta_\mathrm{eff}(t) = -\frac {\ln P(t) - \ln P(t') } {\ln(t) - \ln(t')} \ ,
\end{equation}
using $t - t'=8 \ \mathrm{ or } \ 4$.
These difference selections have been found to be optimal in noise
reduction versus effective exponent range \cite{rmp}.
Similarly, one can also 
define $\eta_\mathrm{eff}(t)$.

\begin{figure}
  \centering
\includegraphics[width=.5\textwidth]{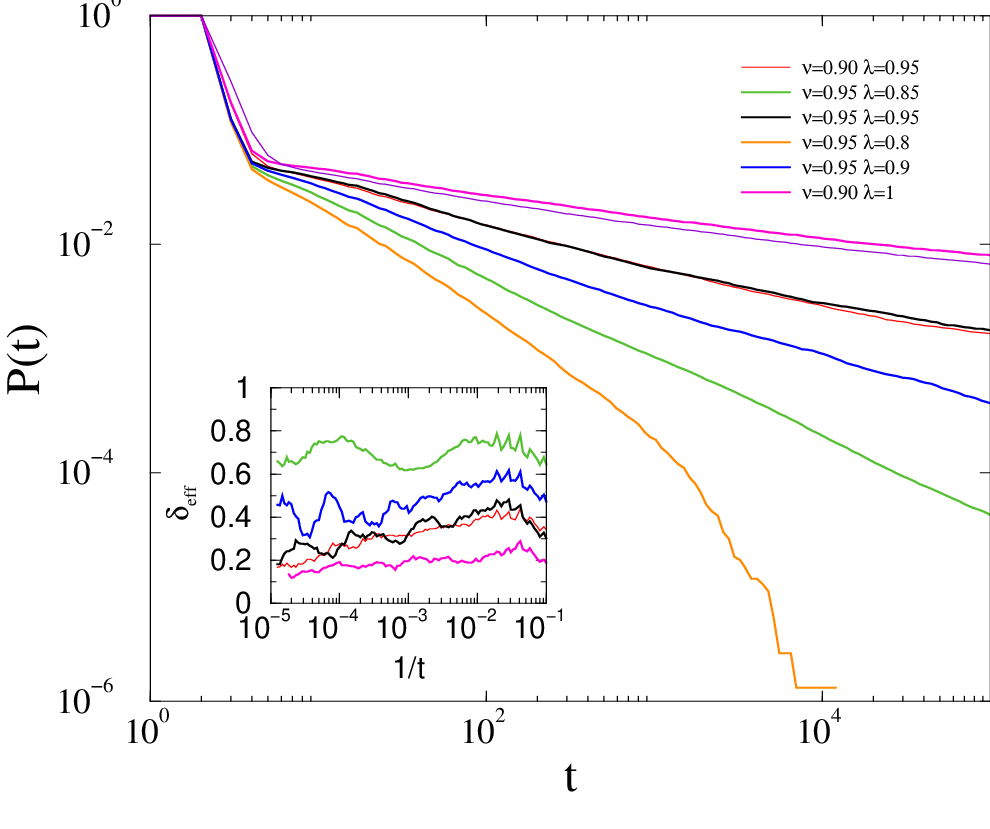}
\caption{\label{ps-0.300.eps}
Avalanche survival probability of the two-state threshold model 
using $K=0.25$, $\nu = 0.90, 0.95$ and 
$\lambda = 0.8$, $0.85$, $0.90$, $0.95$, $1$  
(as shown by the legends).
Inset: Local slopes of the same curves in opposite order
showing saturation of the non-universal exponents in the GP.
Figure from~\cite{GProbcikk}.}
\end{figure}

In  Ref.~\cite{CCdyncikk}, results were obtained for two-state threshold models on directed, 
randomly diluted edge variants of the KKI-18 network.
The results were compared with 
those of the original, undirected graph and showed qualitative invariance 
of the GP for those changes. 
Figure~\ref{ps-0.300.eps} shows the activity avalanche survival
probability $P(t)$ for the threshold $K=0.25$ and 
deactivation probability $\nu = 0.9, 0.95$ on the KKI-18
connectome. One can see power-law tails for more than three decades with
continuously changing exponents as we vary $\lambda$ and $\nu$.

The critical point, above which $P(t)$ signals persistent
activity, is around $\nu_c=0.90(5)$ for $\lambda=1$,
the most efficient activity propagating branching process.
However, it is difficult to locate the critical point exactly because the evolution slows down and 
exhibits oscillating as well as logarithmic corrections.
The decay at the critical point is slower than the mean-field decay,
characterized by $\delta= 0.5$.
It may even be ultra-slow (i.e.\ logarithmic) as in disordered
directed percolation in $d < 4$ 
dimensions~\cite{Vojta2006b}.

Below the transition point, the avalanche survival exponent 
changes continuously in the range $0.2 < \delta < 0.7$ as shown
in the inset of Fig.~\ref{ps-0.300.eps}.
These effective exponents exhibit stabilization of 
$\delta_{\rm eff}(t)$ for $10^2 < t < 10^5$. 
Using the scaling relation (\ref{tau-del}), we get 
dynamical scaling exponents in the GP region: $1.2 < \tau_t \le 1.7$,
overlapping with the human brain experimental values 
$1.5 < \tau < 2.4$ of~\cite{brainexp}. 

\begin{figure}[ht]
  \centering
\includegraphics[width=.5\textwidth]{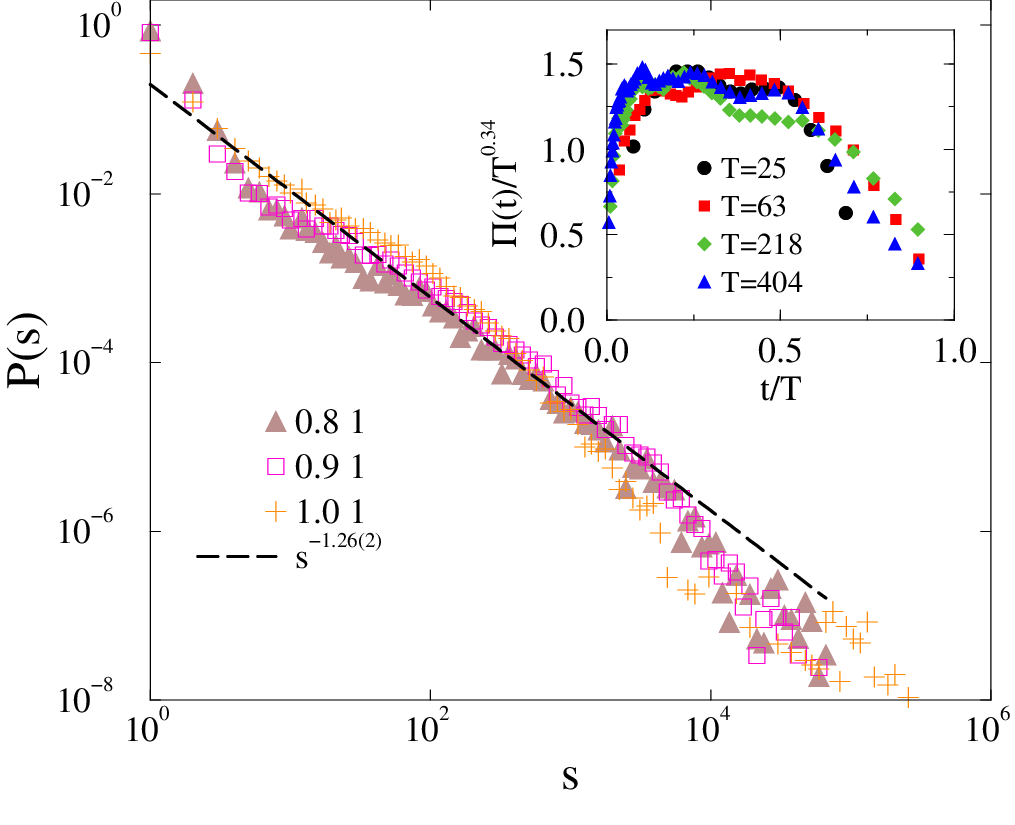}
\caption{\label{elo-t2wlsw} Avalanche size distribution of the two-state 
threshold model at $K=0.25$, $\nu=1$ and $\lambda=1,0.9,0.8$. 
Dashed line: power-law fit to the $\lambda=0.8$ case.
Inset: Avalanche shape collapse for $T= 25, 63, 218, 404$ at
$\lambda=0.86$ and $\nu = 0.95$. Figure from~\cite{CCdyncikk}}
\end{figure}

At and below the critical point, the avalanche size distributions
also exhibit non-universal power-law tails, as shown in Fig.~\ref{elo-t2wlsw}, 
characterized by $\tau > 1.25(2)$, overlapping with the experiments
$1 < \tau < 1.6$ of~\cite{brainexp}. Note, that log-periodic oscillations
are superimposed on the power-laws, as a consequence of the modular
graph topology, thus even in the last decade of the simulation data
$10^4 <  s \le 10^5$ we may see an upbending, instead of a cutoff.
The collapse of averaged avalanche distributions $\Pi(t)$ for fixed 
temporal sizes $T$ as in \cite{Fried} was also studied. The inset of
Fig.~\ref{elo-t2wlsw} displays a good a collapse, obtained for avalanches
of temporal sizes $T = 25, 63, 218, 404$, using a vertical scaling
$\Pi(t)/T^{0.34}$, which is near the experimental findings
reported in~\cite{Fried}. Note also the asymmetric shape which
is in agreement with the experiments but could not be
reproduced by the model of Ref.~\cite{Fried}.

In Ref.~\cite{CCdyncikk}, the connectome modifications included random removal 
of up to $20\%$ of directed connections and flipping the signs of the weights $W_{ij}$ 
of randomly selected edges.
The random link removal affects the long-range connections more,
as they are less frequent, than the short-range ones.
Thus, this modification makes the connectome closer to reality in the
sense of statistics, as it introduces asymmetry in the connections
in an expected fraction and compensates the distortion of the 
MRI diffusion tensor imaging method, which underestimates local fiber tracts~\cite{Tractrev}.
However, this edge thinning made the dynamics slower and
the exponents a little bit smaller than the experimental values, suggesting
that additional factors should also be taken into account.

In Ref.~\cite{GProbcikk}, the two-state model study was extended to a 
three-state, refractory threshold version as well as to a time-dependent 
threshold version with a binary distribution $\{K, K-\Delta K\}$. 
Numerical evidence was provided for the robustness of the GP for both 
modifications. However, the GP shrank if the amplitude of time dependence
was stronger because, for large $\Delta K$, the system could jump over
the control-parameter region where the GP occurs.

\begin{figure}
\centering
\includegraphics[width=.5\textwidth]{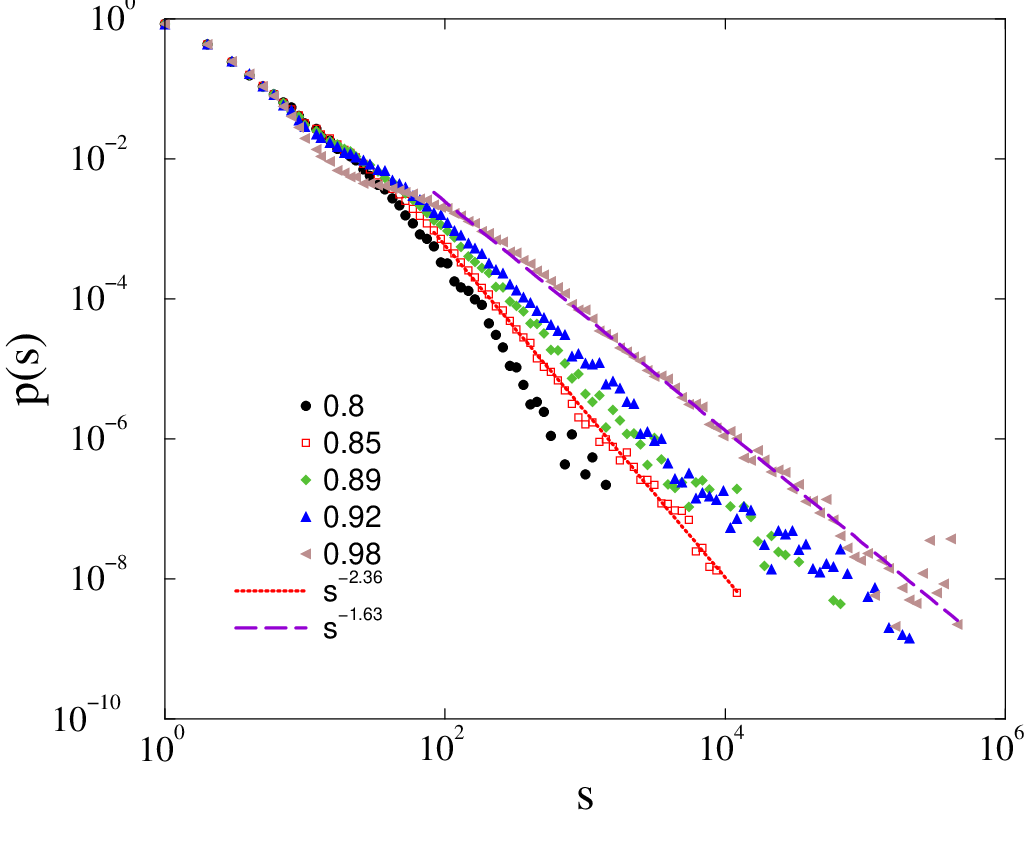}
\caption{\label{elo-TwlsWri}
Avalanche size distributions in the sub-critical phase of the
three-state threshold model with $5\%$ inhibitory links at $K=0.2$, 
$\nu = 1$ and $\lambda = 0.8$, $0.85$, $0.89$, $0.92$, $0.94$, $0.98$ 
(bottom to top curves). Dashed lines show power-law fits for the tails of the 
$\lambda = 0.85$ and the $\lambda = 0.98$ curves with $\tau = 1.63(1)$
and $\tau = 2.36(2)$ respectively. 
}
\end{figure}

This robustness was studied with and without the presence of negative-weighted
edges. Figure~\ref{elo-TwlsWri} shows refractory model results for the avalanche 
sizes in case of $5\%$ randomly flipped inhibitory links at $K=0.2$. 
Non-universal scaling is evident in the GP 
sub-critically for $0.8 < \nu \le 0.98$, characterized by exponent values 
$1.63 \le \tau \le 2.63$.
These values are close to the characteristic times of human 
experiments~\cite{brainexp}: $1 < \tau < 1.6$.
The model results do not change much if inhibitory links are absent~\cite{GProbcikk}.

\begin{figure}
\centering
 \includegraphics[width=.5\textwidth]{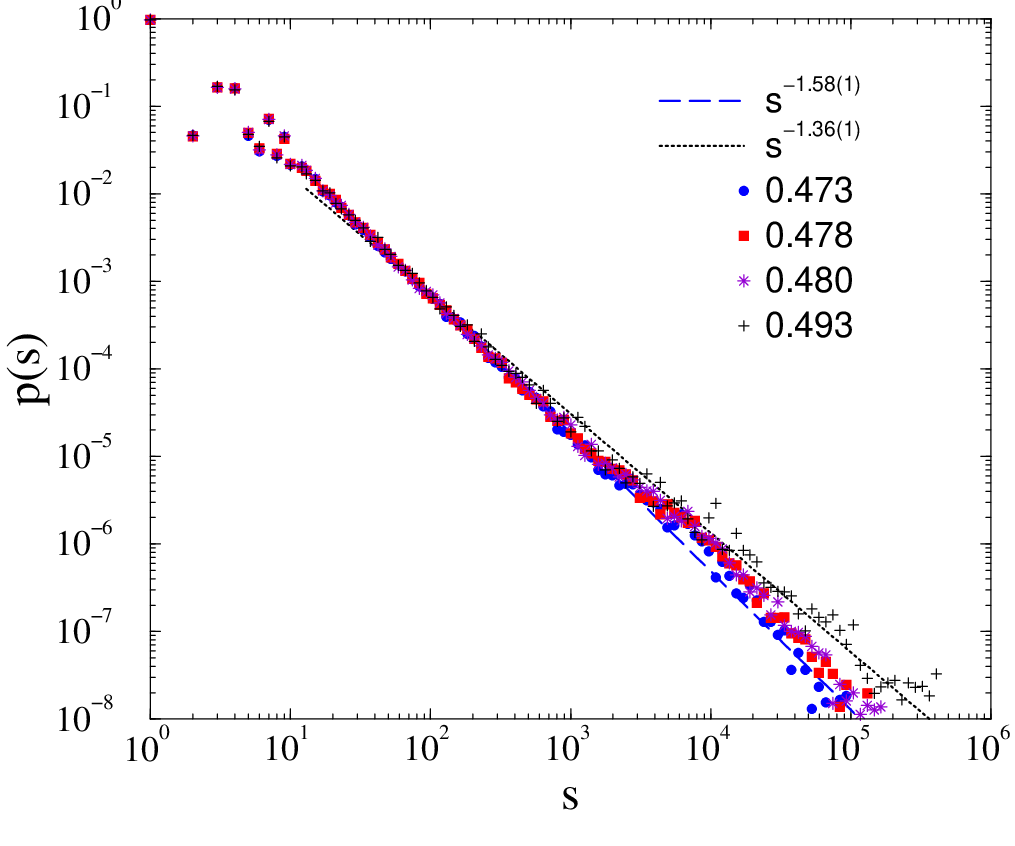}
\caption{\label{ps-0.300-09.eps}
Avalanche size distributions of the two-state threshold model with $30\%$
inhibitory links using time varying threshold: $K=0.1$, $\Delta K=0.01$,
at $\nu = 0.95$ and  $\lambda = 0.473$, $0.478$, $0.480$, $0.483$, $0.493$,
(bottom to top curves).
Dashed lines show power-law fits $s^{-1.58(1)}$ and $s^{-1.36(1)}$ for the 
tails of the $\lambda = 0.493$ and $\lambda = 0.473$ curves. 
Figure from~\cite{GProbcikk}.}
\end{figure}

Figure~\ref{ps-0.300-09.eps} shows some results for the time-dependent two-state model, where the
threshold $K=0.1$ was lowered to $K=0.09$ at randomly selected time steps 
with probability $0.5$. In this case, the GP shrank approximately to the region 
$0.483 < \lambda < 0.513$. Thus, the critical point
moved down to $\lambda  \simeq 0.51(1)$ with respect to the time independent
model.
Therefore, the avalanche size exponent in the GP varies as
$1.36(1) < \tau < 1.58(1)$, well inside the experimental range 
$1 < \tau < 1.6$ of~\cite{brainexp}. 

\section{Critical synchronization dynamics on the Connectome}

One of the most fundamental models showing phase synchronization is the
Kuramoto model of interacting oscillators~\cite{kura}. It was originally
defined on full graphs, corresponding to mean-field
behaviour~\cite{chate_prl}. The critical dynamical behaviour has recently
been explored on random graphs~\cite{cmk2016,Kurcikk}.
A phase transition in the Kuramoto model can happen only above the lower
critical dimension $d_l=4$ \cite{HPClett}.
In lower dimensions, a true, singular phase
transition in the $N\to\infty$ limit is not possible, but partial
synchronization can emerge with a smooth crossover if
oscillators are strongly coupled.

The Kuramoto model describes interacting oscillators with phases~$\theta_i(t)$
located at $N$ nodes of a network, which evolve according to the
dynamical equation
\be
\dot{\theta_i}(t) = \omega_{i,0} + K \sum_{j} W_{ij} \sin[ \theta_j(t)- \theta_i(t) ]
+ s \xi_i(t).
\label{kureq}
\ee
Here, summation is performed over neighbouring nodes of $i$.
$\omega_{i,0}$ is the intrinsic frequency of the $i$-th oscillator, 
drawn from a $g(\omega_{i,0})$ distribution.
For distributions with flat top the transition becomes
discontinuous~\cite{PhysRevE.72.046211,KKIdeco}. Usually, a Gaussian
distribution with zero mean and unit variance is used to study a
continuous synchronization transition.
Oscillatory behaviour is possible in the presence of quenched 
$g(\omega_{i,0})$ self-frequencies.
In their absence,
Eq.~(\ref{kureq}) describes a nonequilibrium XY model (e.g.\ see ~\cite{rmp}).
We can also add an annealed noise $\xi_i(t)$ process, to emulate thermal 
fluctuations, which is a Gaussian white noise in general,
coupled by the amplitude $s$~\cite{KKIdeco}.

The global coupling $K$ is the control parameter of this model by which 
we can tune the system between asynchronous and synchronous states. 
One usually follows the synchronization transition through studying 
the Kuramoto order parameter defined by
\be
R(t)=\frac{1}{N}\left|\sum_{j=1}^Ne^{i\theta_j(t)}\right|,
\label{op}
\ee
which is non-zero above a critical coupling strength $K > K_c$ or tends to
zero for $K < K_c$ as $R \propto\sqrt{1/N}$. At $K_c$, $R$ exhibits growth
as
\be
R(t,N) = N^{-1/2}\, t^{\eta}\, f_{\uparrow}(t / N^{\tilde z}) \ ,
\label{escal}
\ee
with the dynamical exponents ${\tilde z}$ and $\eta$,
if the initial state is incoherent.
Otherwise, the initial
state decays as
\be
R(t,N) = t^{-\delta}\, f_{\downarrow}(t / N^{\tilde z}) \ ,
\label{dscal}
\ee
characterized by the dynamical exponent $\delta$. Here $f_{\uparrow}$ and
$f_{\downarrow}$ denote different scaling functions.

The (noiseless) Kuramoto equation exhibits Galilean symmetry~\cite{pikovsky,Acebron}.
It is invariant to the global shift of a mean rotation frame
$\overline\omega \to \overline\omega'$ and the oscillation-size dependence 
can also be gauged out by the following transformation: 
$\omega_i \to a \omega_i'$, $t \to (1/a)t'$ and $K\to a K'$.
Therefore, for small values of $a$, necessary to transform a normal 
Gaussian $\omega_i$ distribution with $\sigma=1$, to another Gaussian 
with $\sigma=0.02$, corresponding to real empirical data,
we can obtain the same results as for $\sigma=1$ at late times
and small global couplings~\cite{KKIdeco}. 
This scale invariance is an important technical benefit,
which can be exploited to simulate ultra-slow oscillations at time scales
$<0.01\,\mathrm{Hz}$, shown by human-brain fMRI 
measurements~\cite{Ponce-Deco-Plos15,DKJR}.

\begin{figure}[!h]
\centering
\includegraphics[width=.5\textwidth]{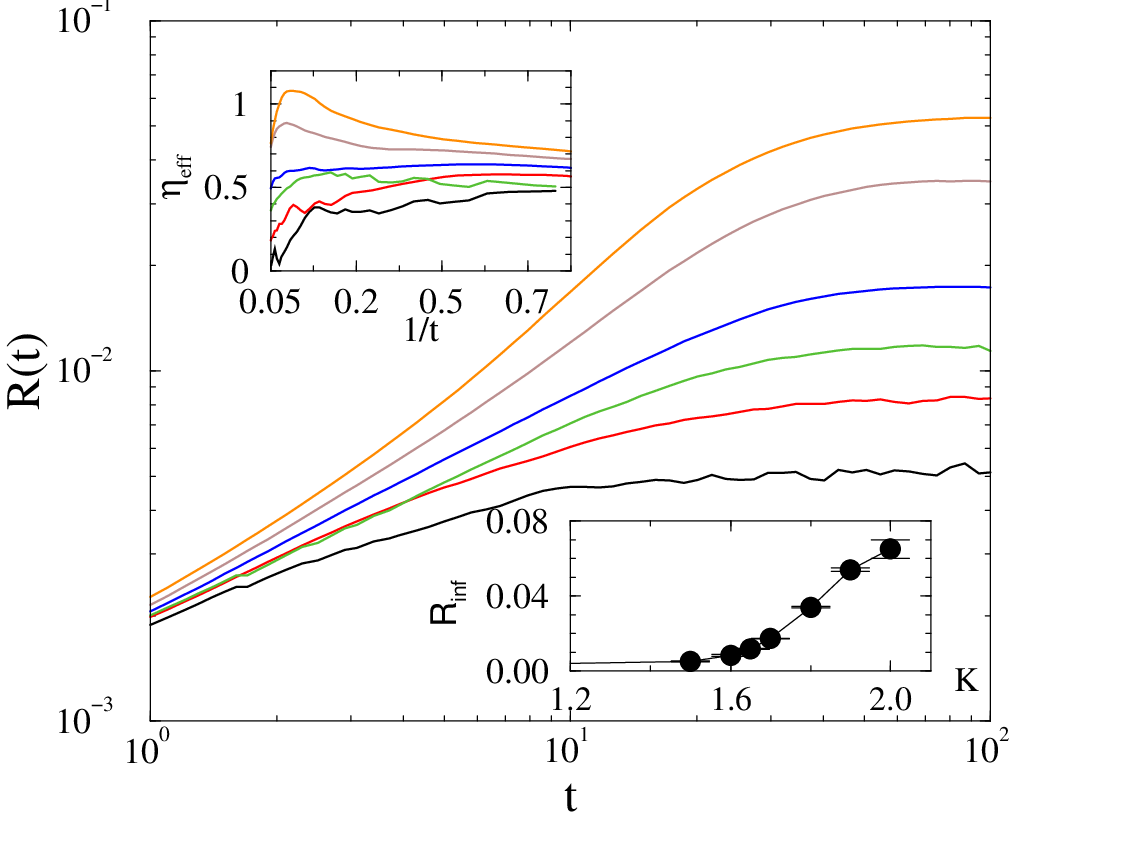}
\caption{Growth of the average $R$ on the KKI-18 graph near the synchronization
transition point for $K=1.5$, $1.6$, $1,65$, $1.7$, $1.8$, $1.9$ (bottom to top curves).
Upper left inset: Effective exponents, defined as (\ref{deff}) for the same data,
down inset: steady state $R(t\to\infty)$ as the function of the global coupling.
Figure from~\cite{KurCC}}
\label{figccgrowth}
\end{figure}

To locate the transition from de-synchronized to synchronized states, one can 
increase the global coupling $K$ and determine $R(t)$ by averaging over 
thousands of realizations with different, independent initial conditions. 
The computer experiments are done by applying random initialization of 
phases with uniform distribution $\theta_i(0) \in (0,2\pi ]$.
Using the parallelized Runge-Kutta-4 algorithm for NVIDIA graphic cards (GPU), 
a 40-fold increase in the throughput could be achieved with respect 
to a single 12-core CPU.
Figure~\ref{figccgrowth} shows an example for the growth of the Kuramoto order parameter
on the KKI-18 connectome.
The crossover is smooth, but the transition point can be estimated visually via the
inflexion condition, which separates up (convex) and down (concave) bending curves
for times $t< 20$ before finite size causes saturation of $R(t)$.
Looking at the local slopes, we can estimate this crossover at $K_c = 1.60(5)$,
with an effective scaling exponent $\eta_\mathrm{eff} \simeq 0.6(1)$.
This is smaller than the $\eta = 0.75$ mean-field value of the Kuramoto model~\cite{cmk2016}.
The lower inset in Fig.~\ref{figccgrowth} shows that the steady state values $R(t\to\infty,K)$ 
exhibit a low level of synchronization even above the transition point.

\begin{figure}[!h]
\centering
\includegraphics[width=.5\textwidth]{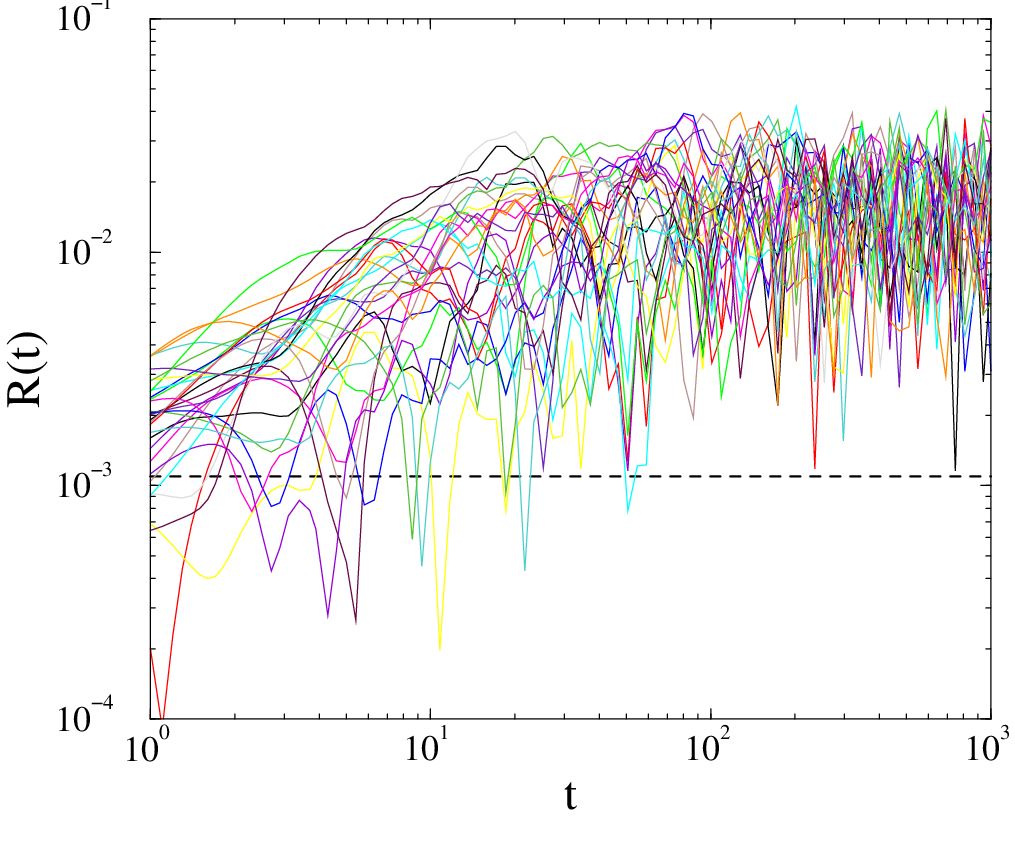}
\caption{Evolution of $R(t)$ for many single realizations on the 
KKI-18 graph at $K = 1.7$.
The dashed line shows the threshold value $R = 1/\sqrt{N} = 0.001094$,
used for the characteristic time measurements. 
Figure from~\cite{KurCC}.}
\label{figR}
\end{figure}

To define synchronization "avalanches" in terms of the Kuramoto order parameter,
we can consider processes, starting from fully de-synchronized initial states by
a single phase perturbation (or by an external phase shift at a node), followed 
by growth and return to $R(t_x) = 1/\sqrt{N}$, corresponding to the disordered
state of $N$ oscillators. In the simulations one can measure the first return, 
crossing times $t_x$ in many random realizations of the system.
In ~\cite{KurCC,KKIdeco}, the return time was estimated by $t_x = (t_k + t_{k-1})/2$, 
where $t_k$ was the first measured crossing time, when $R$ fell below
$1/\sqrt{N} = 0.001094$, see Fig.~\ref{figR}, which is just a demonstration, 
showing the evolution of independent realizations, slightly above the estimated 
transition point. These independent sample evolutions and their average
at a critical point follow the so called initial slip phenomena (see for
example~\cite{odorbook}), followed by a fallback to the disordered state, 
providing a possibility to estimate critical exponents.

\begin{figure}[!h]
\centering
\includegraphics[width=.5\textwidth]{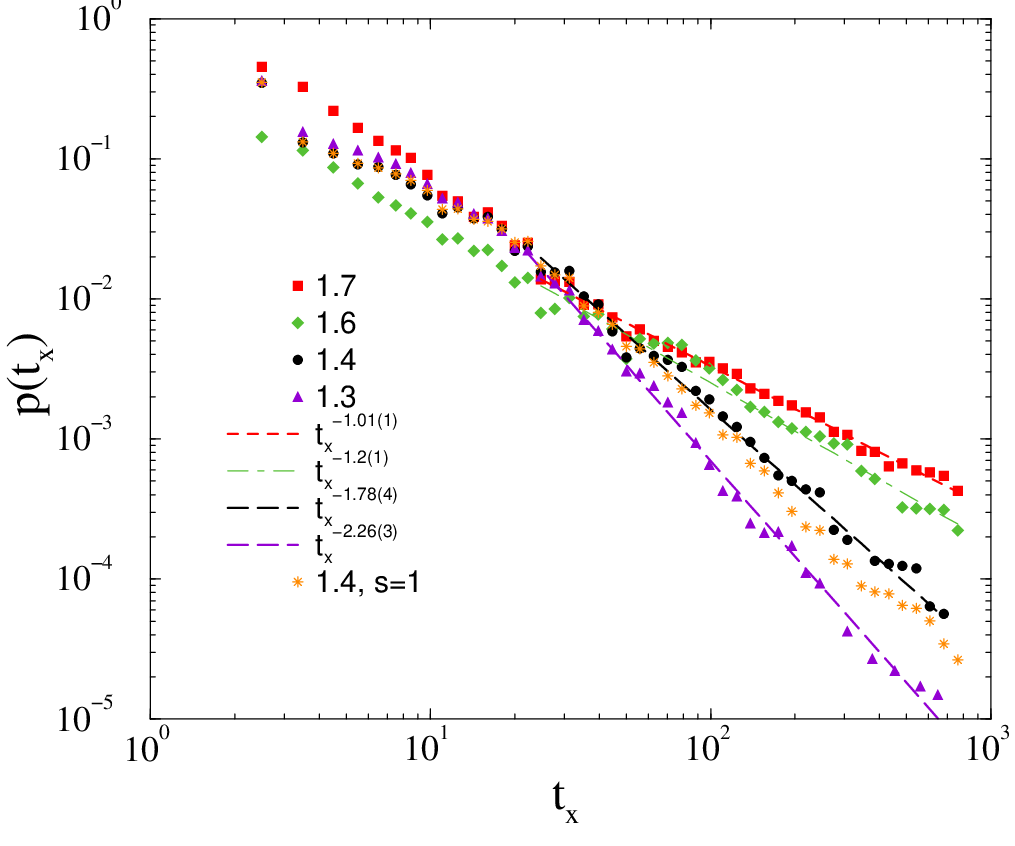}
\caption{Duration distribution of $t_x$ on the KKI-18 model for growth
$K=1.7$ (boxes), $1.6$ (diamonds), $1.4$ (bullets),  $1.3$ (up triangles), 
$1.4$ with $s=1$ noise (asterisks).
The dashed lines shows power-law fits for the tail region $t_x > 20$.
Figure from~\cite{KKIdeco}}
\label{durcc}
\end{figure}

Following a histogramming procedure, one can obtain distributions of
$p(t_x)$, which exhibit power-law tails for $1.2 < K \le 1.7$,
characterized by the exponents $1 < \tau_t < 2$ (see Fig.~\ref{durcc}), 
which are in the range of the in vivo human neuro-experiments: 
$1.5 < \tau_t < 2.4$ of~\cite{brainexp}.
At $K = 1.6$ (i.e.\ near the transition point), we find
$\tau_t=1.2(1)$.
Above
the transition point, the decay
$p(t_x) \sim 1/t$ marks a synchronized phase, where return to 
de-synchronization can take long.
The exponent value $\tau_t=1.2(1)$ at the transition suggests that the real 
brain works in the sub-critical phase, where we can still observe
dynamical criticality. The phase with the non-universal power laws
is reminiscent of GPs, but modules create frustrated 
synchronization regions~\cite{Frus,Frus-noise,FrusB} 
with meta-stability and chimera-like states~\cite{chimera}.
For comparison, on a large two-dimensional lattice with
additional random, long-range links, the mean-field value
$\tau_t \simeq 1.6(1)$ was obtained~\cite{Kurcikk}.

The addition of weak noise does not change these results as can be
seen for $s=1$ at $K=1.4$ in Figure~\ref{durcc}. However, stronger noise
causes deviations, which are difficult to investigate as numerical
precision of the applied Runge-Kutta-4 integration breaks down in
case of large differences that are generated by strong fluctuations
and produced by annealed noise.

Additionally, when the signs of the weights on a randomly selected $5\%$ of
links are flipped as $W''_{ij} = -W'_{ij}$,
dynamical scaling was found to be invariant.
Such links suppress local synchronization and can thus be considered as 
an inhibition mechanism.
The crossover to synchronization occurs at $K_c = 1.9(1)$, slightly higher
than for the original KKI-18 network.
The tails of the $p(t_x)$ probability distributions exhibit power laws with
$1 < \tau_t \le 2$ in the $1.4 < K < 1.8$ region.

In \cite{CCdyncikk}, the robustness of the GP with threshold-model 
dynamical behaviour was tested by randomly neglecting 20\% of 
links in one direction. In~\cite{KurCC}, the neglect of 
all links in one direction ($W''_{ij} = -W'_{ij}$,  $W''_{ji} = 0$)
was investigated. Even in this extremely an-isotropic situation, an 
extended scaling region emerges below the smooth transition point.

\begin{figure}[!h]
\centering
\includegraphics[width=.5\textwidth]{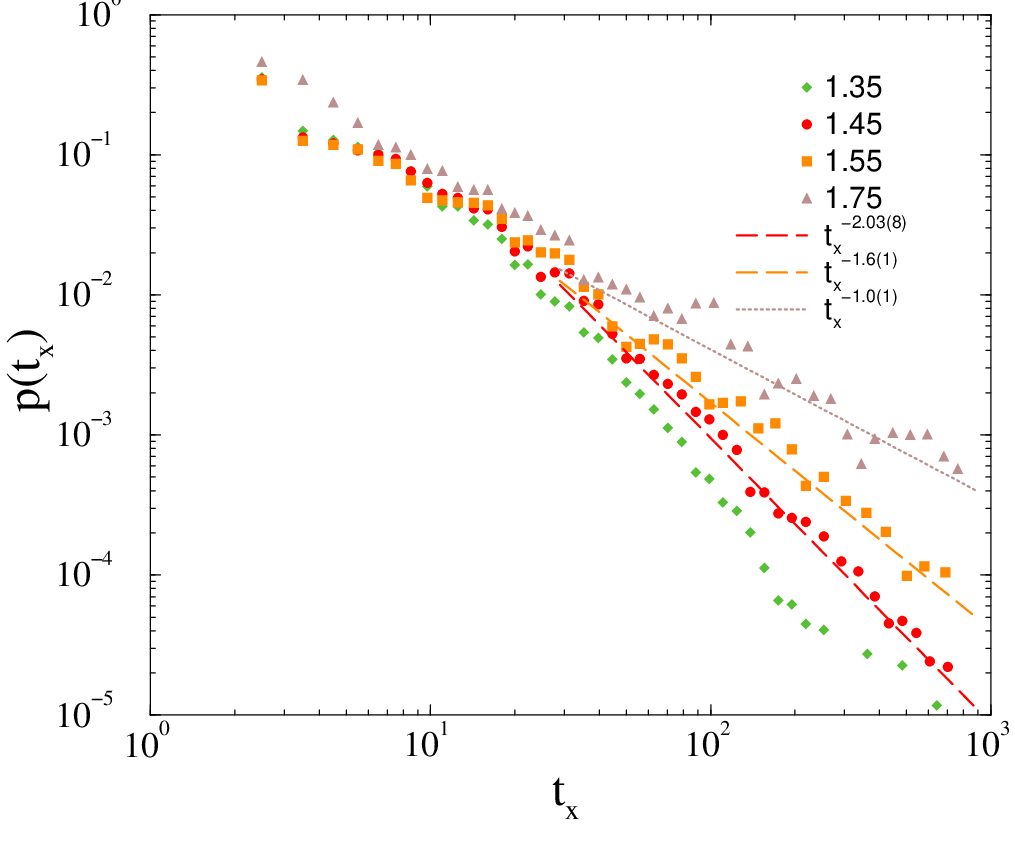}
\caption{Duration distribution of $t_x$ on the KKI-18 model
if $5\%$ of the nodes are inhibitory and $K=1.35$ (diamonds),
$K=1.45$ (bullets), $1.55$ (boxes), $1.75$ (triangles).
The dashed line shows power-law fits to the tail region: $t_x > 20$.
Figure from~\cite{KurCC}}.
\label{figInode5}
\end{figure}

Finally, graphs with $5\%$, $10\%$ and $20\%$ inhibitory nodes were
created by flipping the signs of (outward or inward) link weights of these
randomly selected sites. 
Below the synchronization transition point, which is at $K_c=1.7(1)$
for $5\%$, we can find again a frustrated synchronization region,
where power-law-tailed de-synchronization durations occur as before
(see Fig.~\ref{figInode5}).
The exponent values are in the range $1 < \tau_t < 2$.

\section{Conclusions and Outlook}

Neural variability makes the brain more efficient~\cite{Orb16}, and
one must, therefore, consider its effect in modeling.
To study this effect, extended dynamical simulations have been performed on 
large human connectome models.
Weight heterogeneity of such structural graphs is too strong to allow critical 
behaviour to appear. Note, that the "weights" of the connectome edges are coming 
from MRI techniques and estimate the number of tracks between regions of interest.
But, the real strength or the type (excitatory or inhibitory) of the node
interactions is not encoded in the graphs we could access.
Thus, weight normalized versions were considered, which model local homeostasis
by synaptic inhibition mechanisms.

Inhibitions were modeled by interactions with negative couplings among
nodes. We found that there is no qualitative difference between the
inhibitory link and node models, Griffiths Phases with non-universal
exponents close to experiments were detected in both cases. This is
not surprising from the point of view of statistical physics, but it is also
known that inhibitory neurons are not homogeneously distributed across the
cortex. There is a correlation between inhibition and connectivity.
For example, inhibition often serves to balance excitation in
specific circuits~\cite{Deneve2016}. 
However, this is far from being understood at the whole-brain 
level, except for the hippocampus~\cite{Hainmueller2020}.

After comparing available human brain connectomes using network topology analysis,
one network, called KKI-18, was selected as a representative. For 
the threshold model, critical exponents in the range $1.4 < \tau_t < 1.7$ and
$1.5 < \tau < 2$, close to neural experimental values~\cite{BP03}, were found.

Regarding the oscillatory Kuramoto model, we conclude that
quenched disorder in the self-frequencies causes power-law tails
in the dynamical behaviour of chimera-like states at the edge of criticality. 
These non-universal power laws resemble Griffiths Phase effects.
The scaling laws also resemble results for the second-order
Kuramoto model on power grids below the synchronization
transition~\cite{POWcikk}. We found characteristic dynamical 
time exponents $1 < \tau_t < 2$, overlapping with LRTC experiments.

A recent analysis~\cite{CC18} on new and publicly available 
data from both anesthetized and freely moving animals concluded 
that, if the cortex demands both extreme modes of operation 
(synchronized and de-synchronized) for different functions, it 
may be advantageous to self-organize near and hover over the 
critical point between the two modes. Note, however, that the same 
group reported lately that their results were artifacts of 
sampling~\cite{Carval21}.
Our large scale simulation results show that the characteristic 
time exponents of~\cite{BP03} and of~\cite{BP03} can be reproduced 
with the assumption of GP sub-critically.

It is important to note, that while some rough tuning
of the control parameters might be necessary to get
closer to the critical point, one can see dynamical
criticality even {\it below a phase transition point}
without external activation,
which is a safe expectation for brain systems~\cite{Pris}.
Recent experiments suggest slightly sub-critical brain
states in vivo, devoid of dangerous over-activity linked to
epilepsy.

The dynamical scaling behaviour has been found to be robust, supporting
universality. Although the Kuramoto model could be considered too
simplistic to describe the brain; in the weak-coupling limit quantitative
agreement was found among various classes of oscillators: integrate-and-fire, 
Winfree, and Kuramoto-Daido type for a fully connected network of 
identical units~\cite{PolRos15}. Assuming this holds for heterogenous
models in the sub-critical region, this would provide support for 
the edge-of-criticality hypothesis for oscillating systems near and 
below the synchronization transition point.
Additive weak annealed noise, added to the Kuramoto equation also resulted in
invariance of the scaling~\cite{KKIdeco}.
Gaussian noises with amplitudes not larger than those of the quenched 
Gaussian self-frequencies do not affect the
previous results within numerical precision. This means that
time-dependent, thermal-like noise does not destroy or alter the
dynamical scaling behaviour of this model.
We also pointed out that the empirical results with ultra-slow 
oscillations can be transformed onto zero-mean Gaussian frequencies 
as a consequence of the Galilean symmetry of the Kuramoto 
equation~\cite{KKIdeco}.
Positiveness of the $g(\omega_{i})$ distribution is necessary in
the brain, as we do not expect neural oscillators to 'rotate backwards'.
This corresponds to the question of an asymmetric distribution of natural
frequencies, such that for $g(\omega_{i}) = 0$ for $\omega_{i} < 0$.
It has been shown that, in the case of uni-modal $g(\omega_{i})$-s,
only the first derivative, the flatness of $g(\omega_{i})$, matters.
Without a flat top, like an asymmetric triangle, one obtains the
same universal critical behaviour ($\beta=1/2$) as for the original Kuramoto
model with zero-centered symmetric Gaussian~\cite{BU08}. Thus we expect the same dynamical
behaviour for an asymmetric, truncated Gaussian with $g(\omega_{i}) = 0$ for
$\omega_{i} < 0$.

An interesting continuation could be the study of
the effect of phase shifts, caused by finite signal propagation
in the neural network { or the introduction of a threshold, as in
integrate-and-fire models, although by universality of critical
systems we do not expect qualitative changes in the scaling behaviour.}

Although these connectomes do not provide a true network of the brain, as
for example the nodes themselves are built from thousands of neurons and
may not map the gray matter links well,
they can still lead to the best meso-level approximation for critical
brain dynamics. Further research is under way to extend our approaches
to large, exact, but still not full connectomes available at present.

Sub-sampling ambiguities may also cause differences from the experiments.
Our mesoscopic model could also open up the possibility to clarify this
with the possibility of changing the scale of averaging of simulation
results.

Meta-stability and hysteresis are also common in brain behaviour. 
They are related to the ability to sustain stimulus-selective
persistent activity for working memory~\cite{bistab-exp}.
The brain rapidly switches from one state to another in response to
stimulus, and it may remain in the same state for a long time after
the end of the stimulus. Meta-stability and hysteresis occur in 
general at first-order phase transitions. However, at
hybrid type or mixed order transitions dynamical criticality can
coexist with them. Using synthetic hierarchical modular networks, it has recently been
shown that even GPs can be found below the discontinuous transition 
using threshold type models, where the excitation level are high enough
to cause fragmentation of the possible activity
patterns~\cite{HPTcikk}. In brain science language, this means that
the structural and the functional networks are different.
This allows GPs in high dimensional, small-world graphs, which for
simpler systems was hypothesized to be impossible, leaving out only
mean-field like behaviour~\cite{Munoz2010}. An interesting direction
would be to extend such model analysis using real connectome graphs.

Finally, the mechanistic studies reviewed here offer a causal account of the
role of synchronization and specifically of heterogeneity for criticality at the
whole-brain level. This mechanistic framework is extremely promising and
relevant not only to deepen our understanding of healthy brain functions but also
for its breakdown in neuropsychiatric diseases. Our whole-brain model
perspective might help to improve the diagnosis, and design of therapies after
understanding the subtle synchronization effects relevant in mental diseases.

\section*{Acknowledgments}

G.\'O is supported by the National Research, Development and Innovation 
Office NKFIH under Grant No. K128989 and the Project HPC-EUROPA3 
(INFRAIA-2016-1-730897) from the EC Research Innovation Action under the 
H2020 Programme.

M.T.G. was supported by the Singapore Ministry of Education (MOE) and Yale-NUS College 
(through Grant No. R-607-263-043-121).

G.D. is supported by Spanish national research projects (ref. PID2019-105772GB-I00
MCIU AEI) funded by the Spanish Ministry of Science, Innovation and Universities
(MCIU), State Research Agency (AEI); HBP SGA3 Human Brain Project Specific Grant
Agreement 3 (grant agreement no. 945539), funded by the EU H2020 FET Flagship
programme; SGR Research Support Group support (ref. 2017 SGR 1545), funded by the
Catalan Agency for Management of University and Research Grants (AGAUR); Neurotwin
Digital twins for model-driven non-invasive electrical brain stimulation (grant
agreement ID: 101017716) funded by the EU H2020 FET Proactive programme; euSNN
European School of Network Neuroscience (grant agreement ID: 860563) funded by the
EU H2020 MSCA-ITN Innovative Training Networks; CECH The Emerging Human Brain
Cluster (Id. 001-P-001682) within the framework of the European Research Development
Fund Operational Program of Catalonia 2014-2020; Brain-Connects: Brain Connectivity
during Stroke Recovery and Rehabilitation (id. 201725.33) funded by the Fundacio La
Marato TV3;  Corticity, FLAG˙˙ERA JTC 2017, (ref. PCI2018-092891) funded by the
Spanish Ministry of Science, Innovation and Universities (MCIU), State Research
Agency (AEI).

J. K. is supported by the Helmholtz Initiative and Networking Funds via the 
W2/W3 programme, project number W2/W3-026.

We gratefully acknowledge computational resources provided by the
Hungarian National Supercomputer Network, the BSC Barcelona and the 
HZDR computing center.

\section*{References}
\bibliography{paper}

\end{document}